\documentclass[12pt]{article}

\usepackage[dvips]{graphicx}
\usepackage{epsfig,cite}
\usepackage{amsmath}
\usepackage{color}
\usepackage{amssymb}
\usepackage{slashed}
\usepackage{jheppub}
\usepackage{hyperref} 
\include{epsf}
\usepackage[normalem]{ulem}

\newcount\timecount
\newcount\hours \newcount\minutes  \newcount\temp \newcount\pmhours
\hours = \time
\divide\hours by 60
\temp = \hours
\multiply\temp by 60
\minutes = \time
\advance\minutes by -\temp
\def\hour{\the\hours}
\def\minute{\ifnum\minutes<10 0\the\minutes
            \else\the\minutes\fi}
\def\clock{
\ifnum\hours=0 12:\minute\ AM
\else\ifnum\hours<12 \hour:\minute\ AM
      \else\ifnum\hours=12 12:\minute\ PM
            \else\ifnum\hours>12
                 \pmhours=\hours
                 \advance\pmhours by -12
                 \the\pmhours:\minute\ PM
                 \fi
            \fi
      \fi
\fi
}

\def\monthname{\relax\ifcase\month 0/\or January\or February\or
   March\or April\or May\or June\or July\or August\or September\or
   October\or November\or December\else\number\month/\fi}

\def\bold#1{\setbox0=\hbox{$#1$}%
     \kern-.025em\copy0\kern-\wd0
     \kern.05em\copy0\kern-\wd0
     \kern-.025em\raise.0433em\box0 }

\def\beq{\begin{equation}}
\def\eeq{\end{equation}}

\def\ga{\mathrel{\raise.3ex\hbox{$>$\kern-.75em\lower1ex\hbox{$\sim$}}}}
\def\la{\mathrel{\raise.3ex\hbox{$<$\kern-.75em\lower1ex\hbox{$\sim$}}}}
\def\gev{{\rm \, Ge\kern-0.125em V}}
\def\tev{{\rm \, Te\kern-0.125em V}}
\def\gyr{{\rm \, G\kern-0.125em yr}}

\def\gappeq{\mathrel{\rlap {\raise.5ex\hbox{$>$}}
{\lower.5ex\hbox{$\sim$}}}}
\def\lappeq{\mathrel{\rlap{\raise.5ex\hbox{$<$}}
{\lower.5ex\hbox{$\sim$}}}}
\def\Toprel#1\over#2{\mathrel{\mathop{#2}\limits^{#1}}}

\def\m12{m_{1\!/2}}

\def\bea{\begin{eqnarray}}
\def\eea{\end{eqnarray}}
\def\beqn{\begin{eqnarray}}
\def\eeqn{\end{eqnarray}}

\def\beqar{\begin{eqnarray}}
\def\eeqar{\end{eqnarray}}

\begin{document}

\begin{titlepage}
\pagestyle{empty}
\baselineskip=21pt
\rightline{KCL-PH-TH/2016-13, LCTS/2016-08}
\rightline{CERN-TH/2016-064, IPPP/16/24}
\vskip 0.8in
\begin{center}
{\large {\bf Search for Sphalerons: IceCube vs. LHC}}

\end{center}
\begin{center}
\vskip 0.4in
 {\bf John~Ellis}$^{1,2}$
{\bf Kazuki~Sakurai}$^3$
and {\bf Michael~Spannowsky}$^3$
\vskip 0.1in
{\small {\it
$^1${Theoretical Particle Physics and Cosmology Group, Physics Department, \\
King's College London, London WC2R 2LS, UK}\\
\vskip 0.1in
$^2${Theoretical Physics Department, CERN, CH-1211 Geneva 23, Switzerland}\\
\vskip 0.1in
$^3${Institute for Particle Physics Phenomenology, 
Department of Physics, University of Durham, Science Laboratories, South Road, Durham, DH1 3LE, UK}\\
}}
\vskip 0.6in
{\bf Abstract}
\end{center}
\baselineskip=18pt \noindent
{\small
We discuss the observability of neutrino-induced sphaleron transitions in the IceCube detector,
encouraged by a recent paper by Tye and Wong (TW), which argued on the basis of
a Bloch wave function in the periodic sphaleron potential that such transitions
should be enhanced compared to most previous calculations. We calculate the dependence on
neutrino energy of the sphaleron transition rate, comparing it to that for conventional
neutrino interactions, and we discuss the observability of tau and multi-muon production in
sphaleron-induced transitions. We use IceCube 4-year data to constrain the sphaleron rate,
finding that it is comparable to the upper limit inferred previously from a recast of an ATLAS search for
microscopic black holes at the LHC with $\sim 3$/fb of collisions at 13~TeV.
The IceCube constraint is stronger for a sphaleron barrier height $E_{\rm Sph} \gtrsim 9$~TeV,
and would be comparable with the prospective LHC sensitivity with 300/fb of data at 14~TeV
if $E_{\rm Sph} \sim 11$~TeV.
}


\vfill
\leftline{
March 2016}
\end{titlepage}
\baselineskip=18pt
 
\section{Introduction}
The recent discovery of the Higgs boson with a mass of $125$ GeV by ATLAS and CMS \cite{Aad:2012tfa, Chatrchyan:2012xdj} completes the electroweak sector of the Standard Model. Ongoing measurements of interactions of Higgs and gauge bosons find good agreement with perturbative predictions of the Standard Model  with its spontaneously broken $SU(2) \times U(1)$ gauge group \cite{LEP-2,ATLAS-CONF-2015-044,ATLAS:2012mec,Aad:2012awa,CMS:2014xja,Chatrchyan:2014bza}. A direct consequence of the $SU(2)$ gauge group, beyond the perturbative regime, is the existence of non-perturbative topological effects. 
Field configurations with finite Euclidean action are classified by an integer topological winding number, namely the Chern-Simons number $N_\mathrm{CS}$. Topologically distinct ground states are separated by an energy barrier, and the sphaleron \cite{Tye:2015tva} is an extremal saddle point on top of the barrier with half-integer $N_\mathrm{CS}$ and an energy $E_\mathrm{Sph} \simeq 9$ TeV. Owing to the Adler-Bell-Jackiw anomaly, transitions through sphaleron configurations would lead to striking (B+L)-violating processes. Direct observation of such interactions can be of capital importance in explaining the mechanism underlying the cosmological baryon asymmetry \cite{Kuzmin:1985mm, Fukugita:1986hr, Rubakov:1996vz,Cohen:1993nk,Buchmuller:2004nz}, which might arise from the transmutation of a primordial lepton asymmetry.

While the energy of the sphaleron is directly linked to the shape of the effective potential for the Chern-Simons number and is thus rather undisputed, its production rate is subject to large theoretical uncertainties \cite{McLerran:1989ab, Khoze:1990bm, Ringwald:2003ns}. Over the years, there have been many estimates of the rate of sphaleron transitions in high-energy collisions,
most of them with discouraging results for the prospects for experimental searches. However,  a new approach \cite{Tye:2015tva}, exploiting the periodicity of the Chern-Simons potential, has challenged this paradigm and argued in favour of enhanced sphaleron-induced transition rates.

The authors indicate that, since the effective Chern-Simons potential is periodic, it is
appropriate to use a Bloch wave function to estimate the sphaleron-induced
transition rate. They find that, although the transition rate is still strongly suppressed at
energies below the sphaleron threshold energy, the rate may not be suppressed at higher energies above $E_{\rm Sph}$.

Motivated by this suggestion, two of us (JE and KS) have analyzed the observability of sphaleron-induced
transitions at the LHC and possible future higher-energy $pp$ colliders, and recast a
recent ATLAS search for microscopic black holes using $\sim 3$/fb of data at 13~TeV
in the centre-of-mass as a search for sphalerons \cite{Ellis:2016ast}. In this way, expanding on previous proposals \cite{Gibbs:1995bt, Ringwald:2002sw}, we were able to establish for the first time a significant direct experimental constraint on sphaleron-induced transitions,
assess the potential improvement in sensitivity of future LHC runs with higher integrated 
luminosities and possibly energy, and preview the corresponding prospects for possible
future higher-energy $pp$ colliders.

In this connection, it is important to remember that cosmic rays provide collisions with
centre-of-mass energies beyond those attainable with the LHC. However, the low
fraction of ultra-high-energy $pp$ collisions that would produce sphaleron transitions,
combination with the limitations of the Auger experiment~\cite{deMelloNeto20141476} for extracting detailed information from air showers renders challenging this avenue in searches for sphalerons \cite{Brooijmans:2016lfv}.

Better prospects may be offered by ultra-high-energy neutrino events \cite{Morris:1993wg, Ringwald:2001vk, Fodor:2003bn, Han:2003ru,Illana:2004qc}, which could be observed in a cubic-kilometer neutrino telescopes, such as IceCube \cite{Ahrens:2003ix, icecube_DD} and KM3NeT \cite{Coniglione:2015aqa}.
As we discuss in more detail below, the estimates of TW suggest that the rate of
sphaleron-induced transitions in the highest-energy cosmic-ray collisions observed by
IceCube could be comparable to the conventional neutrino cross section, and might
even dominate the interactions of cosmogenic neutrinos \cite{Beresinsky:1969qj} produced by the Greisen-Zatsepin-Kuzmin
(GZK) process \cite{Greisen:1966jv, Zatsepin:1966jv}: $p + \gamma_{\rm CMB} \to \pi^+ \to \nu$. 

Accordingly, in this paper we extend the analysis of ES to ultra-high-energy 
neutrino interactions, estimating the upper limit on sphaleron transitions obtainable 
from present IceCube results and considering the implications for GZK neutrinos.
Remarkably, we find that the present IceCube sensitivity is very similar to that of
the first LHC data at 13~TeV. If the rate of sphaleron transitions were to saturate
the present LHC limit, the rate of GZK neutrino interactions would be significantly
higher than is conventionally estimated, improving significantly the prospects for
their future detection with IceCube or a cubic-kilometre detector.

\section{Neutrino-Nucleon Cross-Section Calculations}

The analysis of~\cite{Tye:2015tva} is based on the idea that sphaleron transitions 
changing the Chern-Simons number $n$ can be modelled by
considering a Bloch wave function for an effective one-dimensional
Schr\"odinger equation
\begin{equation}
\left( - \frac{1}{2 m} \frac{\partial^2}{\partial Q^2} + V(Q) \right) \Psi(Q) \; = \; E \Psi(Q) \, ,
\label{schrod}
\end{equation}
where $m$ is an effective ``mass" parameter variously estimated to be $\sim 17.1$~TeV~\cite{Manton:1983nd}
to $\sim 22.5$~TeV~\cite{Akiba:1988ay} and the effective potential is taken from~\cite{Manton:1983nd}:
\begin{equation}
V(Q) \; \simeq \; 4.75 \left(1.31 \sin^2 (Q m_W) + 0.60 \sin^4 (Q m_W) \right) \, {\rm TeV} \, .
\label{VQ}
\end{equation}
The sphaleron barrier height $E_{\rm Sph}$ is the maximum value of the effective potential $V(Q)$,
which is $E_{\rm Sph} = 9.11$~TeV in a pure $SU(2)$ theory, and is estimated to be reduced by $\sim 1$\%
when the $U(1)$ of the Standard Model is included. Following~\cite{Tye:2015tva},
we assume $E_{\rm Sph} = 9$~TeV as a nominal value, but present some numerical results for
$E_{\rm Sph} \in [8, 11]$~TeV. 

Ref.~\cite{Tye:2015tva} found that the pass-band structure in the Bloch wave function approach
reproduced the expected (near-exponential) tunnelling suppression of sphaleron transitions in collisions of quark partons
with subprocess centre-of-mass energies $\sqrt{\hat s} \ll E_{\rm Sph}$, but found that there is no suppression for 
$\sqrt{\hat s} \ge E_{\rm Sph}$.
They found that the rates of sphaleron-induced transitions were similar for the two values of $m$, and
adopted $m = 17.1$~TeV for definiteness.
The result of the TW analysis can be expressed as partonic cross-section 
\begin{equation}
\hat \sigma ( \Delta n = \pm 1 ) \; \propto \; \exp \left( c \frac{4 \pi}{\alpha_W} S(\sqrt{\hat s}) \right) \, ,
\label{sigmaTW}
\end{equation}
where $c \sim 2$ and the suppression factor $S(\sqrt{\hat s})$ is shown in Fig.~8 of~\cite{Tye:2015tva}. 
As discussed in~\cite{Ellis:2016ast}, we approximate $S(\sqrt{\hat s})$ at intermediate energies by
\begin{equation}
S(\sqrt{\hat s}) \; = \; (1 - a) \overline{\sqrt{\hat s}} + a \overline{ {\hat s}} - 1 \; \; \; \; {\rm for} \; \; \; \; 0 \; \le \overline{\sqrt{\hat s}} \; \le \; 1 \, ,
\label{numericalS}
\end{equation}
where $\overline{\sqrt{\hat s}} \equiv \sqrt{\hat s}/E_{\rm Sph}$ and $a = - 0.005$.

In the absence of a reliable calculation of the overall magnitude of Eq.~\eqref{sigmaTW}, following \cite{Ellis:2016ast} we parametrize
the partonic cross section for the sphaleron-induced neutrino-quark collision as  
\beq
\hat \sigma_{q \nu}(\hat s) \; = \; \frac{p}{m_W^2} \, ,
\label{sigma}
\eeq
for $\sqrt{\hat s} > E_{\rm Sph}$ and apply the suppression factor (\ref{numericalS}) for $\sqrt{\hat s} < E_{\rm Sph}$.
Our numerical results are relatively insensitive to the form of this suppression factor.
The overall factor $p$ in (\ref{sigma}) depends in general on $\hat s$~\cite{Tye:2015tva}.
However, our result is also not very sensitive to such an energy dependence, since (as we discuss below)
the interaction is dominated by subprocess energies near the threshold $\sqrt{\hat s} \gtrsim E_{\rm Sph}$,
due to the sharply-falling cosmogenic neutrino flux.
The cross section for sphaleron transitions in neutrino-nucleon collisions is given by
\beq
\sigma_{\nu N} (E_\nu) = \sum_q \int^1_0 dx f_q(x, \mu) \hat \sigma_{q \nu}(2 x m_N E_\nu ) \,,
\eeq  
where $f_q(x, \mu)$ is a parton distribution function for the quark flavour $q$
and $m_N$ is the mass of nucleon.
The neutrino-nucleon centre-of-mass energy $\hat E = \sqrt{2 m_N E_{\nu}}$, neglecting the $m^2_N$ term,
and the neutrino-parton subprocess centre-of-mass energy $\sqrt{\hat s} = \sqrt{2 x m_N E_{\nu}}$.

\begin{figure}[t!]
\begin{center}
\includegraphics[height=8.5cm]{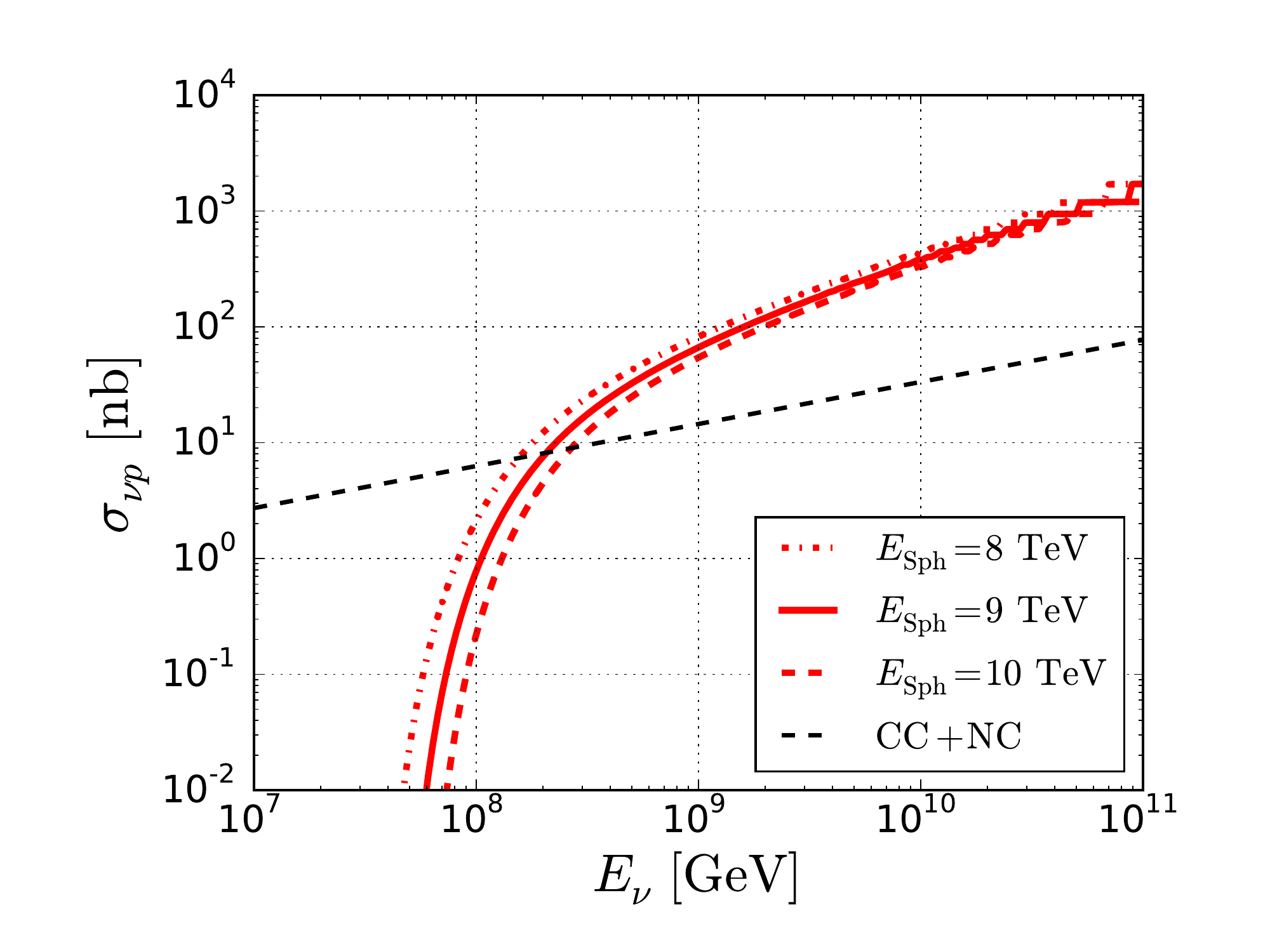}
\end{center}   
\vspace{-0.5cm}
\caption{\label{fig:Edep}\it 
Compared to the sum of the conventional charged- and neutral-current neutrino cross sections (black dashed line),
we show the energy dependence of the cross section for sphaleron transitions in neutrino collisions for a barrier height
$E_{\rm Sph} = 9$~TeV, $c = 2$ and $p = 1$ in Eq.~\eqref{sigma} with
$S$ given by (\protect\ref{numericalS}) (red solid curve) and, for comparison, choices $E_{\rm Sph} = 8$
and $10$~TeV (red dot-dashed and dashed lines, respectively). The variations in the sphaleron curves for $1 \le c \le 4$
are within the widths of the lines, but we recall that the overall normalization factor $p$ is quite uncertain.
}
\end{figure}

Fig.~\ref{fig:Edep} displays the energy dependence of the cross section for sphaleron-induced
transitions calculated under these assumptions with $c = 2$ and $p = 1$: the results are insensitive to $c \in [1, 4]$
and scale linearly with $p$. The solid (dot-dashed) (dashed) red lines are for $E_{\rm Sph} = 9 (8) (10)$~TeV,
and the black dashed line is the sum of the conventional charged- and neutral-current neutrino cross sections.
We see that the sphaleron-induced cross section would dominate for $E_\nu \gtrsim 2 \times 10^8$~TeV
if $p = 1$, but recall that this factor is quite uncertain.

\begin{figure}[t!]
\begin{center}
\includegraphics[height=8.5cm]{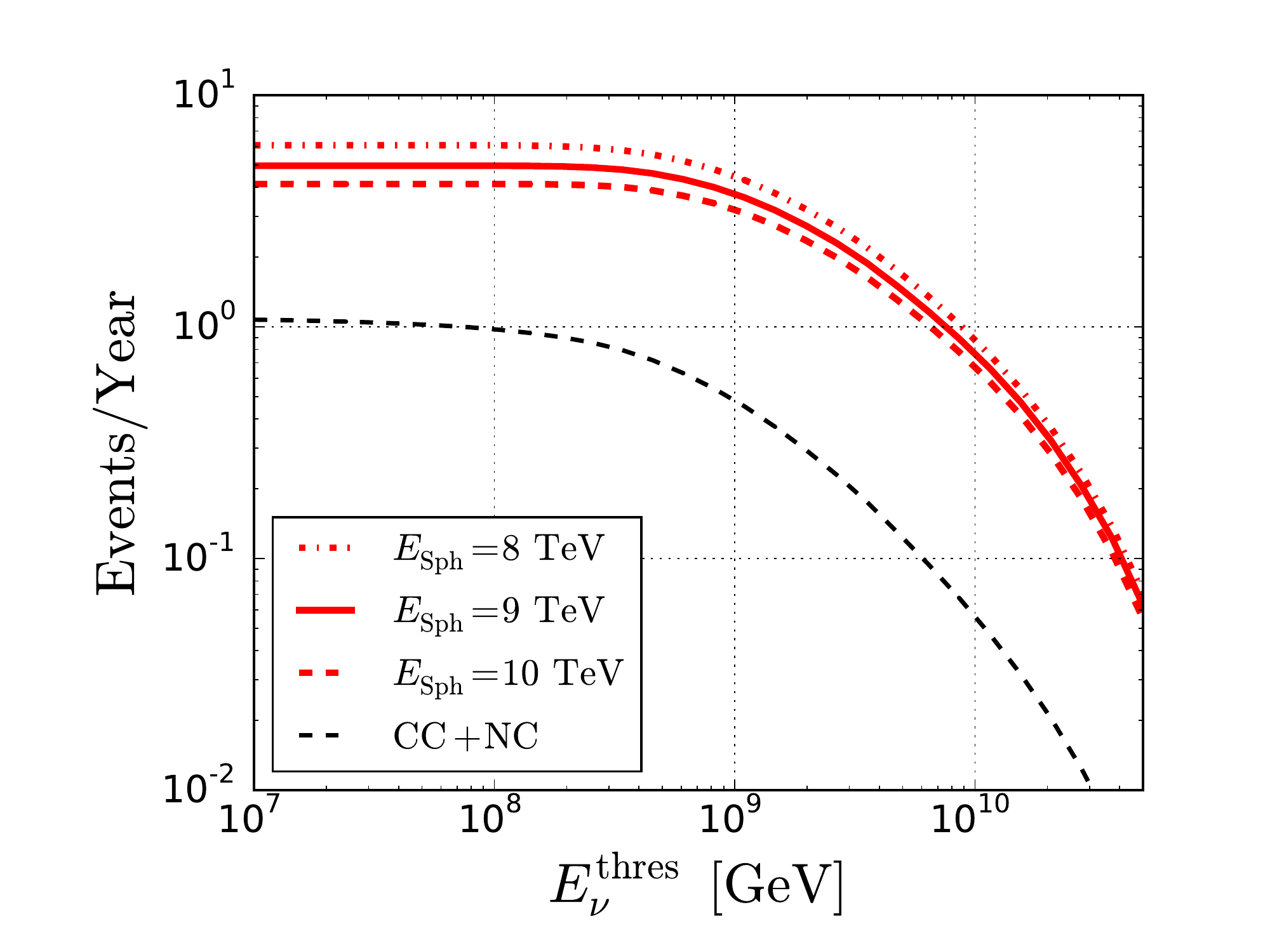}
\end{center}   
\vspace{-0.5cm}
\caption{\label{fig:evrate}\it 
Compared to the sum of the conventional charged- and neutral-current neutrino cross sections (black dashed line),
we show the rate for sphaleron transitions in IceCube for a barrier height
$E_{\rm Sph} = 9$~TeV, $p = 1$ in Eq.~\eqref{sigma} (red solid curve) and, for comparison, choices $E_{\rm Sph} = 8$
and $10$~TeV (red dot-dashed and dashed lines, respectively).
}
\end{figure}

The cross section estimates in Fig.~\ref{fig:Edep} can be convoluted with the cosmogenic neutrino flux, $d^2 \Phi / (d E_\nu d t d \Omega)$ 
[$\rm GeV^{-1} \, \rm cm^{-2} \, s^{-1} \, sr^{-1}$],
to calculate the event rates.
We use the cosmogenic neutrino flux estimated in \cite{Ahlers:2010fw} throughout this paper.
The event rate in the IceCube detector also depends on the energy-dependent effective neutrino detection area, $A_{\rm eff}(E_\nu)$,
which has been evaluated by the IceCube collaboration \cite{Aartsen:2013dsm} using conventional neutrino-nucleon interaction.
Assuming the same detection efficiency, we estimate the sphaleron-induced IceCube event rate as 
\beq
\frac{d N_{\rm Sph}}{d t} = \int_{E_\nu^{\rm thres}} d E_\nu \int d \Omega
\frac{\sigma^{\rm Sph}_{\nu N}(E_\nu)}{ \sigma^{\rm CC/NC}_{\nu N}(E_\nu)} A_{\rm eff}(E_\nu) \frac{d^2 \Phi}{d E_\nu d t d \Omega} \,,
\eeq
where $E_\nu^{\rm thres}$ is the energy threshold of incoming cosmogenic neutrinos.
In the second integral we take into account only neutrinos coming from the upper hemisphere of IceCube,
since the neutrinos from the lower hemisphere will be absorbed by the interaction with the Earth.
In Fig.~\ref{fig:evrate} we show the sphaleron-induced and conventional IceCube event rate as functions of $E_\nu^{\rm thres}$
again assuming $c = 2$ and $p = 1$ and using (dot-dashed) (dashed) red lines
for $E_{\rm Sph} = 9 (8) (10)$~TeV and a black dashed line for sum of the conventional charged- and neutral-current neutrino cross sections.
We see that the sphaleron-induced transitions would dominate over conventional neutrino collisions
by a factor $\gtrsim 5$ for all $E_\nu^{\rm thres} \ge 10^7$~GeV if $p = 1$.


\begin{figure}[t!]
\begin{center}
\includegraphics[height=6cm]{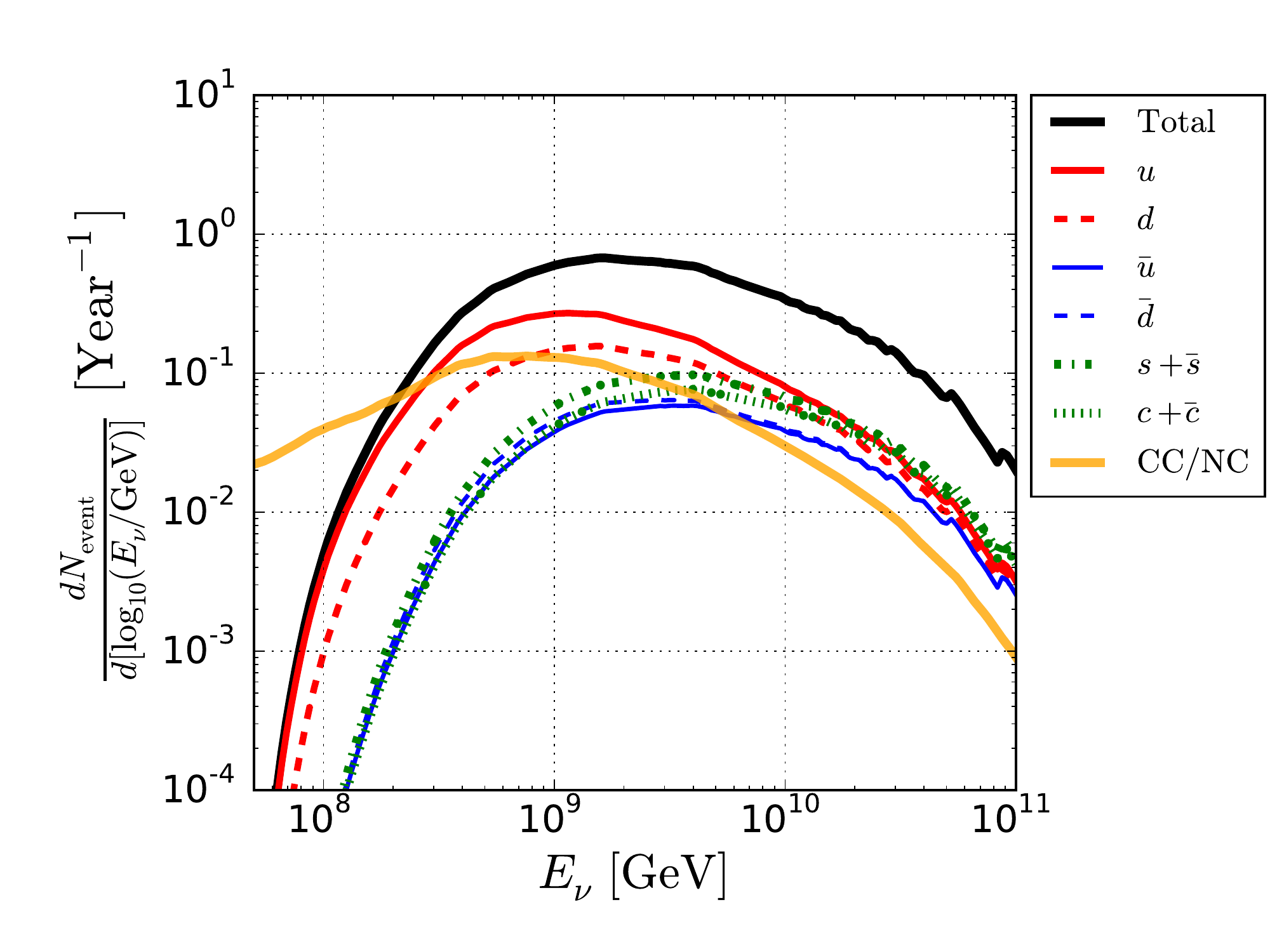} 
\includegraphics[height=6cm]{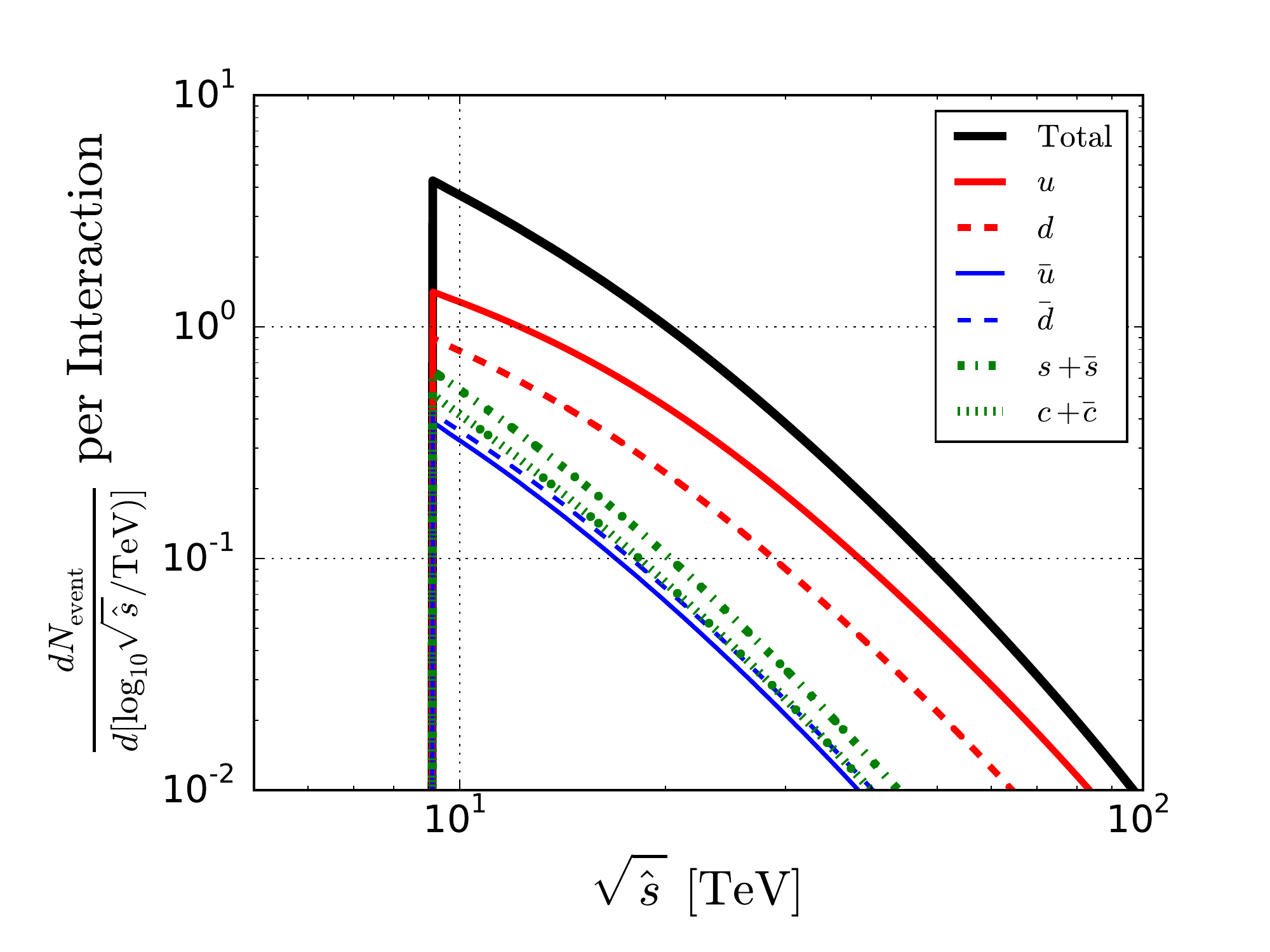} 
\end{center}   
\caption{\label{fig:characteristics}\it 
Left panel: Contributions to the total cross section for sphaleron transitions in neutrino collisions in IceCube, 
for the nominal case $E_{\rm Sph} = 9$~TeV and $p = 1$ in (\protect\ref{sigma}) with
$S$ given by (\protect\ref{numericalS}). The solid green curve is for the sum of conventional
charged- and neutral-current neutrino interactions. Right panel: The distributions in the
neutrino-parton reduced centre-of-mass energy, which is peaked at $E_{\rm Sph}$.
In both panels, the contributions of different parton-parton collision processes are colour-coded
as indicated.
}
\end{figure}

Fig.~\ref{fig:characteristics} displays some characteristics of the sphaleron-induced transitions.
In the left panel we show a breakdown of the collision rates with respect to the quark parton
species inside the nucleon targets in the ice. As was to be expected, interactions with $u$ and $d$ quarks
dominate, followed by interactions with antiquarks and heavy flavours. In the right panel we show
the corresponding distributions in the reduced neutrino-quark subprocess centre-of-mass energies
$\sqrt{\hat s}$, which are sharply peaked at the sphaleron energy $E_{\rm Sph}$, taken here to
have its nominal value of 9~TeV. This peaking implies that our results would not be affected
strongly by a possible energy dependence in the overall factor $p$, but depend essentially only
on the value of $p$ at the sphaleron threshold energy.

\section{Leptons in Sphaleron-Induced Transitions}

In the IceCube detector \cite{icecube_DD},  
neutral current interaction and charged current interaction of electron neutrinos 
leave a shower-like signature,
whilst high energy muons and very high energy taus ($E_\tau > 10^7$ GeV)
leave a track-like signature. 
IceCube expects to be able to see a `double-bang' signature for
$\tau$ leptons with energies $\in [10^6, 10^7]$~GeV. 

We simulate distributions of leptons ($\mu$ and $\tau$) produced by the sphaleron-induced neutrino-quark collision events in parton level. 
We consider the simplest possibility of such events: $q \nu \to 8 \bar q 2 \bar \ell$ induced by
the gauge invariant $(\bar q \bar q \bar q)_1 (\bar q \bar q \bar q)_2 (\bar q \bar q \bar q)_3 (\bar \ell_1 \bar \ell_2 \bar \ell_3)$ operator,
where the suffix denotes the generation.
We assume equal flux for each flavour of cosmogenic neutrinos. 
Leptons can be produced either directly from the {\it primary} interaction, $q \nu \to 8 \bar q 2 \bar \ell$, 
or {\it secondarily} from the decay of the heavy particles ($t$ and $W$).

\begin{figure}[t!]
\begin{center}
\includegraphics[height=6cm]{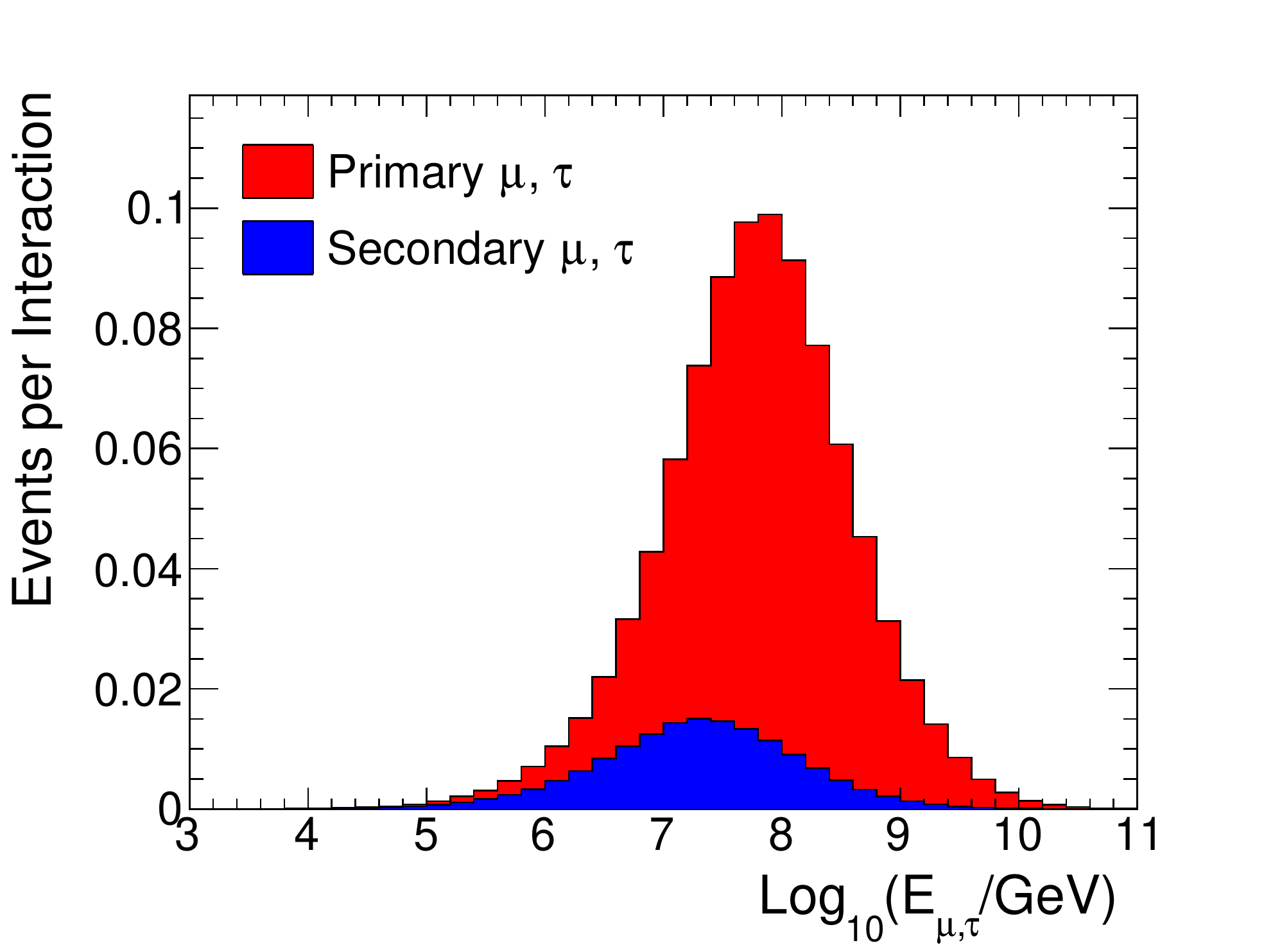} 
\includegraphics[height=6cm]{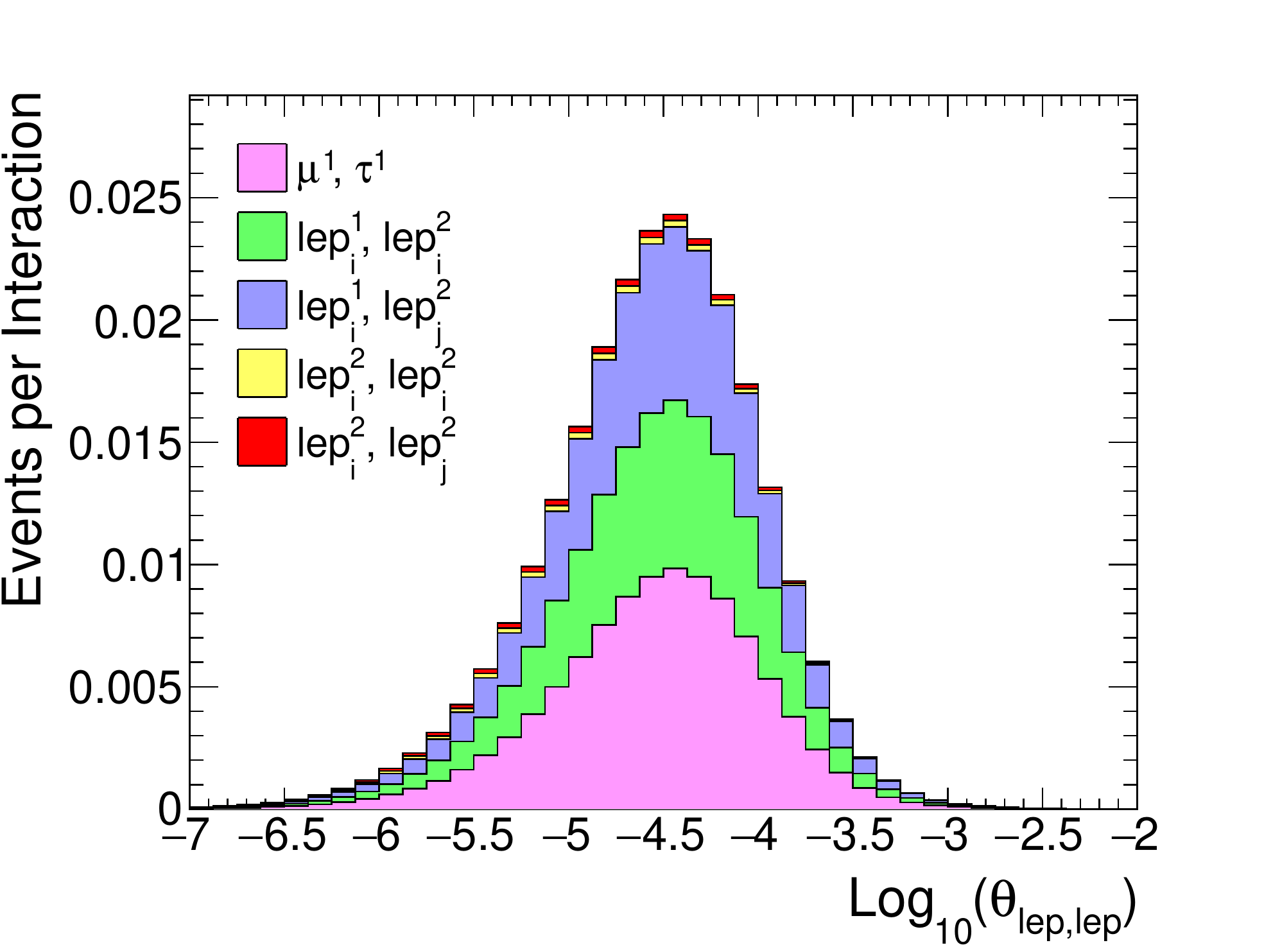} 
\end{center}   
\caption{\label{fig:leptons}\it 
Left panel: Histograms of the primary and secondary muon energy distributions (red and blue, respectively)
in sphaleron-induced transitions in neutrino-nucleon collisions in IceCube for $E_{\rm Sph} = 9$~TeV,
normalised to a single sphaleron-induced event. Right panel: 
Histograms of the opening angles in the laboratory frame between pairs of leptons in sphaleron transitions in neutrino interactions
in IceCube for $E_{\rm Sph} = 9$~TeV, colour-coded for the different combinations of
primary and secondary leptons.
}
\end{figure}

The left panel of Fig.~\ref{fig:leptons} displays the primary and secondary $\mu$ and $\tau$ energy distributions
(which are identical) normalised to a single sphaleron-induced event. We see that the primary lepton energies
are peaked just below $10^8$~GeV, whereas the secondary lepton energies are peaked
closer to $10^7$~GeV. IceCube expects to be able to see a `double-bang' signature for
$\tau$ leptons with energies $\in [10^6, 10^7]$~GeV. We see that sphaleron-induced
transitions would produce some primary and secondary $\tau$ leptons in this energy range.
However, we find only 5\% of the sphaleron-induced events have $\tau$ leptons in this energy range. 

Since the sphaleron-induced interaction can produce multiple leptons, IceCube might be able to see 
multiple tracks in the event if those tracks are separated more than 17\,m.
This separation typically requires the opening angle of the leptons to be $\gtrsim 2 \cdot 10^{-2}$.  
The right panel of Fig.~\ref{fig:leptons} displays histograms of the $\ell -\ell$ 
opening angles in the laboratory frame, colour-coded for the different combinations of
primary and secondary leptons, and again normalised to a single sphaleron-induced event.
We see that the opening angles are in all cases much smaller than the IceCube angular resolution,
so we do not expect multiple lepton tracks to be distinguished.

\section{IceCube Constraints on Sphaleron-Induced Transitions}

\begin{figure}[t!]
\begin{center}
\includegraphics[height=8.7cm]{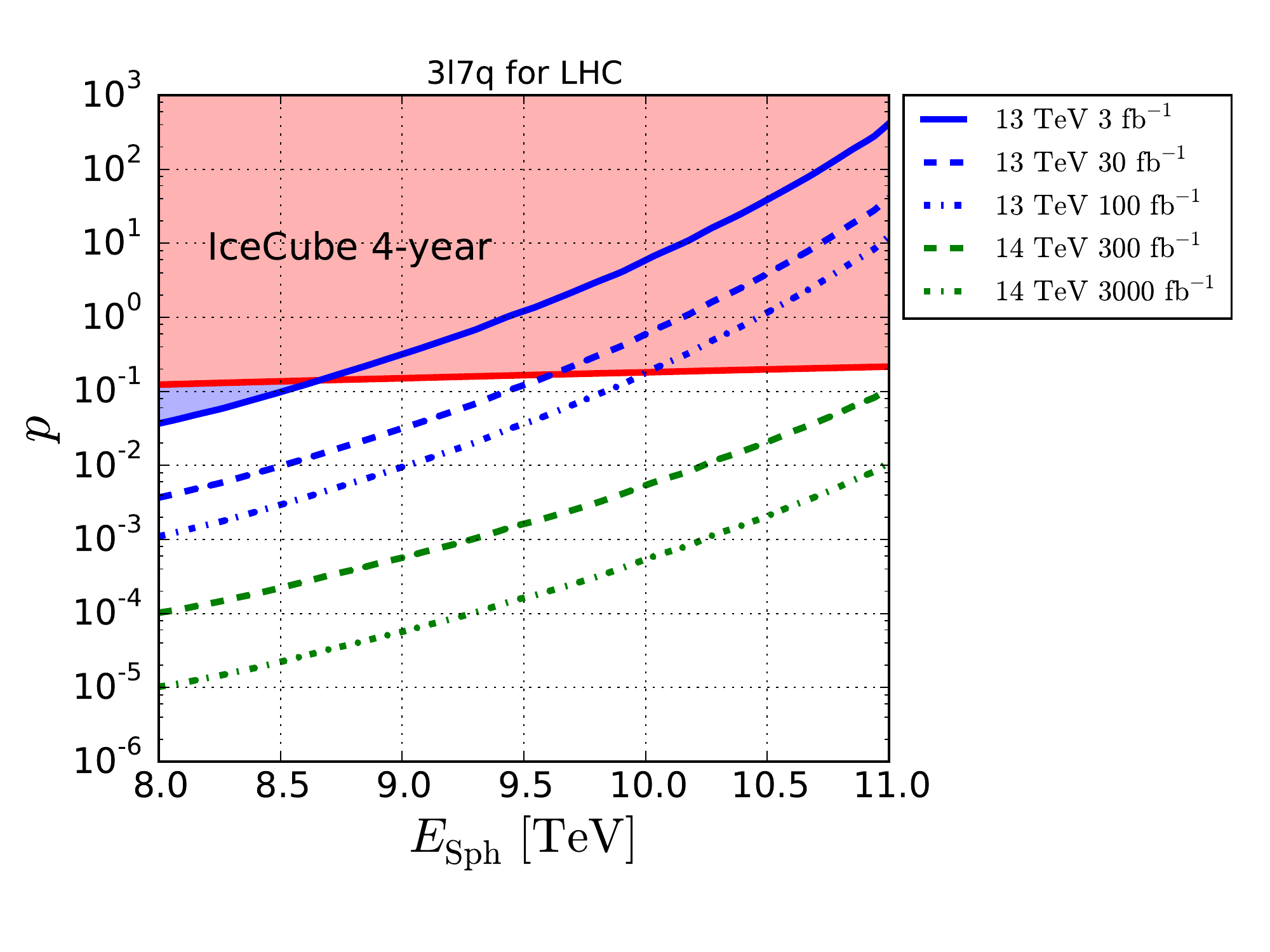} \\
\includegraphics[height=8.7cm]{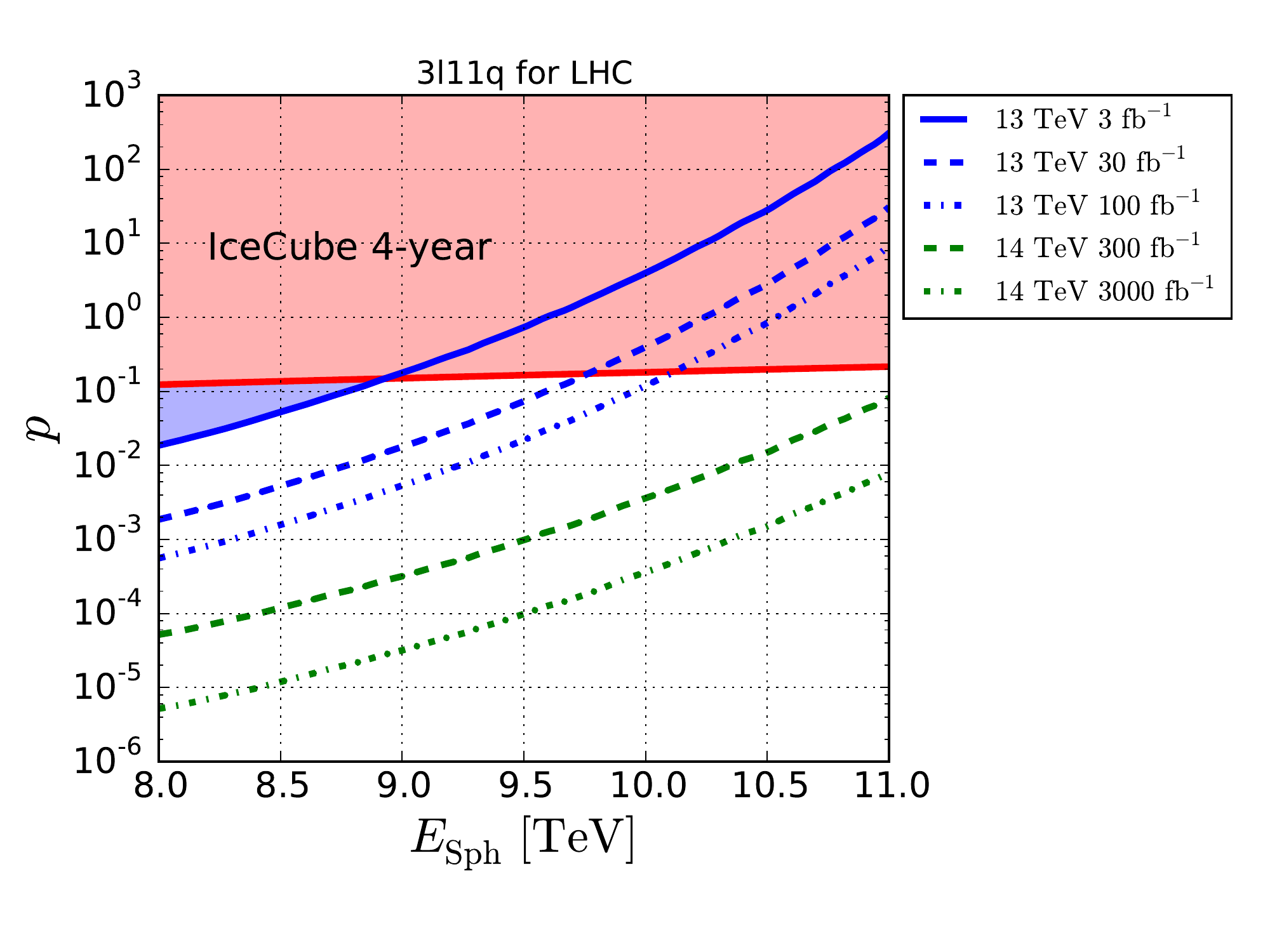} 
\\
\vspace{-0.5cm}
\end{center}   
\caption{\label{fig:limits}\it 
Comparisons of the constraint obtained from IceCube 4-year data \cite{Montaruli:2015six} (near-horizontal red solid line)
with that obtained from a recast of the ATLAS search for microscopic black holes with $\sim 3$/fb
of collisions at 13~TeV (solid blue line). Also shown are prospective LHC sensitivities with
increased luminosity and/or centre-of-mass energy. The comparisons are for $8~{\rm TeV} \le E_{\rm Sph} \le 10~{\rm TeV}$
for $\Delta n = -1$ sphaleron transitions (upper panel) and $\Delta n = +1$ transitions (lower panel).
}
\end{figure}

In the absence of a distinctive leptonic signature, we use the generic IceCube search 
for which detection efficiency is encoded in the effective neutrino detection area given in \cite{Aartsen:2013dsm}.
Moreover, we assume in the absence of a detailed
simulation of the IceCube efficiency for detecting sphaleron-induced final states that it is the same
as that for conventional final states,
and that the neutrino spectrum keeps falling at energies above $10^{11}$ GeV.
Fig.~\ref{fig:limits} compares the upper limit on the
overall cross-section factor $p$ obtained in this way from IceCube 4-year \cite{Montaruli:2015six}
(solid red lines) with the upper limits derived in~\cite{Ellis:2016ast}
from recasting the ATLAS Run~2 search for microscopic black holes with $\sim 3$/fb of data at
13~TeV \cite{Aad:2015mzg} (solid blue lines). The upper panel of Fig.~\ref{fig:limits} is for $\Delta n = -1$ transitions, which yield
final states with 10 energetic particles at the LHC, and the lower panel is for $\Delta n = +1$ transitions, which yield
14-particle final states at the LHC.

The LHC constraints in Fig.~\ref{fig:limits} are different for the 10- and 14-particle final states, and quite
sensitive to the assumed value of the sphaleron energy $E_{\rm Sph}$. (We recall that
our nominal value is $E_{\rm Sph} = 9$~TeV, but we display results for $E_{\rm Sph} \in [8, 11]$~TeV.)
This is because the rate for sphaleron-induced transitions at the LHC with a centre-of-mass
energy of 13~TeV is quite sensitive to $E_{\rm Sph}$. In contrast, the IceCube 4-year limit
is quite insensitive to $E_{\rm Sph}$ over the range studied, because of the larger range of
neutrino energies. Within our assumptions, the IceCube efficiencies and hence limits for $\Delta n = \pm 1$ transitions
are the same, whereas the LHC limits are stronger for $\Delta n = +1$ transitions, for which we
estimated in~\cite{Ellis:2016ast} a greater detection efficiency. 

By a remarkable coincidence, we see that the LHC and IceCube constraints are almost identical
for the nominal value $E_{\rm Sph} = 9$~TeV, but the IceCube limits are stronger for larger
$E_{\rm Sph}$, becoming some 3 orders of magnitude stronger for $E_{\rm Sph} = 11$~TeV.

We also display in Fig.~\ref{fig:limits} the prospective future LHC exclusion sensitivities for higher
integrated luminosities (dashed and dot-dashed blue lines) and energy (dashed and dot-dashed
green lines). We see that 300/fb of luminosity at 14~TeV would be needed for the LHC sensitivity
to surpass the IceCube constraint for $E_{\rm Sph} = 11$~TeV. We anticipate that the IceCube sensitvity
will also be improved by longer operating time and/or effective size, and note that an order-of-magnitude
improvement in the IceCube sensitivity would make it highly competitive with the LHC with 3000/fb for
$E_{\rm Sph} = 11$~TeV, with both being able to reach $p \simeq 10^{-2}$. On the other hand,
for the nominal value $E_{\rm Sph} = 9$~TeV, the LHC would have a greater reach than IceCube, down to
$p < 10^{-4}$.

\section{Summary and Conclusions}

We have shown that IceCube could have a sensitivity to spahaleron-induced transitions
that is comparable to that of the LHC. For a cross-section prefactor $p = 1$, the rate
of such transitions in neutrino collisions would exceed the sum of conventional
charged- and neutral-current interactions for $E_\nu \gtrsim 2 \times 10^8$~GeV,
as seen in Fig.~\ref{fig:Edep}, yielding a larger number of events for a neutrino threshold
energy above $E_\nu = 10^7$~GeV, as seen in Fig.~\ref{fig:evrate}. Our simulations
of neutrino-induced sphaleron transitions do not reveal any distinctive leptonic signatures,
with a limited fraction of `double-bang' $\tau$ events, as seen in the left panel of Fig.~\ref{fig:leptons},
and multilepton bundles that are probably not resolvable, as seen in the right panel of Fig.~\ref{fig:leptons}.

Remarkably, the prospective IceCube constraints on sphaleron-induced transitions are
comparable to those from the LHC, as seen in Fig.~\ref{fig:limits}, with IceCube having an
advantage for large sphaleron energies $E_{\rm Sph}$ and the LHC at small $E_{\rm Sph}$.
The crossover is currently close to the nominal value $E_{\rm Sph} = 9$~TeV.

Our estimates need to be validated by dedicated experimental simulations for IceCube
as well as for the LHC, but our results indicate that both have interesting sensitivities
for sphaleron-induced transitions, able to probe significantly below $p =1$, and hence
able to test or constrain the suggestion by Tye and Wong~\cite{Tye:2015tva} that sphaleron transitions
may be much less suppressed than commonly thought previously.

\section*{Acknowledgements}

The work of JE was supported partly by the London Centre for Terauniverse Studies (LCTS), using funding from the European Research Council via the Advanced Investigator Grant 26732, and partly by the STFC Grant ST/L000326/1. He thanks Henry Tye,
Sam Wong and Andy Cohen for instructive discussions. KS and MS are supported by STFC through the IPPP grant.

\addcontentsline{toc}{part}{Bibliography}
\bibliographystyle{JHEP}
\bibliography{references}


\markboth{}{}

\end{document}